\documentclass[12pt,preprint]{aastex}
\usepackage{graphicx}
\def\gr{$\gamma$-ray }
\def\grs{$\gamma$-rays }
\def\grr{$\gamma$-radiation }
\newcommand{\be}{\begin{eqnarray}}
\newcommand{\ee}{\end{eqnarray}}

\begin{document}
\shorttitle{CR propagation parameters from diffuse VHE \gr emission}
\title{Obtaining cosmic ray propagation parameters from diffuse VHE \gr emission from the Galactic center ridge}

\author{I. B\"usching}
\affil{North-West University, Potchefstroom Campus,  South Africa}
\author{O.C. de Jager}
\affil{North-West University, Potchefstroom Campus,  South Africa}
\author{J. Snyman}
\affil{North-West University, Potchefstroom Campus,  South Africa}
\keywords{diffuse emission, cosmic ray propagation, Galactic center}
\begin{abstract}
The recent discovery of diffuse, VHE $\gamma$ radiation from the Galactic
center ridge by the H.E.S.S. telescope allow for the first time the direct 
determination of  parameters of galactic cosmic ray propagation models.
In this paper we show that the diffuse \grr near the Galactic center may be
explained by the interaction of VHE cosmic ray (CR) protons with the interstellar
gas located in several giant molecular clouds leading to a measurement of
the cosmic ray diffusion coefficient for the galactic center region
of $\kappa = 1.3$ kpc$^2$Myr$^{-1}$ for a mean proton energy of $\sim 3$ TeV, if 
we assume that the CR protons originated from a supernova event (Sgr A East), which took off about 10\,kyr ago. This value of $\kappa$ is $\sim 5$ to 10 times smaller than the locally measured value.

\end{abstract}

\section{Introduction}
The High Energy Stereoscopic System of four telescopes offers currently
the best angular resolution for the study of VHE $\gamma$-rays from cosmic
sources \citep{aharonian06}. With an angular resolution of $\sim 0.08^{\circ}$, the H.E.S.S. Collaboration
was able to resolve $\gamma$-rays associated with molecular clouds in the 
galactic center region \citep{aharonian06}: Whereas a relatively good
correlation was found between the $\gamma$-ray and CS (with the latter measured by \citet{tsuboi99}) surface brightness
distributions within 150 pc ($\ell \sim \pm 1^{\circ}$ along galactic longitude) from the galactic center,
this correlation degraded at a distance of 200 pc from the galactic center (i.e. at $\ell \sim 1.5^{\circ}$).
A strong indicator that this diffuse component is indirectly associated with a source at the
galactic center is the similarity of the spectral indices of the point source HESS\,J1745-290
at the center and this newly discovered diffuse extended emission. Furthermore, the relatively good correlation
within $\ell\sim \pm 1^{\circ}$, but with degrading correlation beyond that,
suggests that we are dealing with a source (e.g. the SNR Sgr A East or the central Black Hole
Sgr A*) at the GC, which was active for
some time in the past and that we are now seeing the  high energetic particles diffusing
from this central source \citep{aharonian06}.  
The most likely primary species responsible for the gamma-ray emission is then
protons, since electrons would have to compete against synchrotron losses,
resulting in a spectral steepening towards large distances. Apart from slight effects
of energy dependent diffusion, protons, in a good approximation, do not loose energy within this environment, 
resulting in an approximate invariant spectral index with distance.

Whereas these results are important from an Astrophysical viewpoint, it is also
of importance from a cosmic ray viewpoint, i.e. the study of cosmic ray propagation
in our galaxy: \citet{aharonian06} suggested that the diffusion coefficient 
for protons in the 4 to 40 TeV range should be less than $10^{30}$ cm$^2$s$^{-1}$
(or 3.5 kpc$^2$Myr$^{-1}$)
as a result of enhanced turbulence and higher magnetic field strengths in the GC region.
The H.E.S.S. data therefore offer a unique possibility to measure the diffusion
coefficient in this part of the galaxy and to compare with other measurements
of the galactic diffusion coefficient. 
This is in particular important as it has been shown \citep{buesching05} that, given 
SNR are the main sources of CR, the widely used method to obtain propagation parameters
by fitting secondary to primary data is at least tainted, as the CR primary component then shows strong
variations in space and time.

In this paper we will model a transient source at the GC with an
activity time-scale in the past. By solving the transport equation
for proton propagation along the galactic plane, we will obtain
the range of diffusion coefficients, which fit the observed HESS profile best.
To do this we will also model the gas distribution as traced by the CS emission.

\citet{tsuboi99}
was the first to obtain a full coverage of the Galactic Center
Bow (GCB) and the molecular cloud structures in CS in the region of interest. Since the
gamma-ray surface brightness is reflected by the line-of-sight integral of the
product of the cosmic ray and gas densities, we will introduce 3D structures
in the GC region (even though not strictly modelling the GCB itself), such that
the line-of-sight integral through these model structures reproduce the true
line-of-sight integrals through the gas density within $\sim 5\%$ accuracy, as described
in more detail below.

Using the diffusive model of CR propagation, together
with observations made by the High Energy Stereoscopic System
(H.E.S.S.) \citep{aharonian06} it is possible to get an
estimation of the diffusion coefficient controlling cosmic ray (CR)
transport in the galactic center region.
\subsection{\grs from pion decay}
The omnidirectional (i.e. integrated over solid angle) differential
\gr source function $q_{\pi^0}(E_{\gamma},\vec{r})$ at the position 
$\vec{r}=(l,b,r)$ 
for the decay $\pi^0\rightarrow 2\gamma$ is given by \citep{buesching01}
\be
q_{\pi^0}(E_{\gamma},\vec{r})&=& 2\int_{\eta}^{\infty} 
\frac{Q_{\pi^0}(\gamma_{\pi},\vec{r})}{\sqrt{\gamma^2_{\pi}-1}}d\gamma_{\pi}
\label{grs:srcfkt}
\ee
where $\gamma_{\pi}$ is the pion Lorentz factor.
The lower boundary of the integration is given by
\be
\eta&=&\frac{E_{\gamma}}{m_{\pi}c^2}+\frac{m_{\pi}c^2}{4\,E_{\gamma}}.
\ee
The pion source function then is 
\be
Q_{\pi^0}(\gamma_{\pi},\vec{r})&=&\rho_{\rm gas}(\vec{r})\,c 
\int_{\gamma_{\rm thr}}^{\infty} \beta\sigma^{\pi^0}_{pp}(\gamma_p,\gamma_{\pi^0})N_p(\gamma_p,\vec{r})d\gamma_p
\label{grs:losi}
\ee
where $\sigma^{\pi^0}_{pp}$ is the total cross section for pion production in $pp$ collisions and $N_p$ the CR proton spectrum.
The differential photon flux from the decay of CR induced $\pi^0$s from the 
direction $(l,b)$ is given by integrating Eq.\ref{grs:srcfkt} along the line of sight 
\be
\frac{dN(E_{\gamma},l,b)}{dt\,dE_{\gamma}\,d\Omega} &=&
\frac{1}{4\pi}\int q_{\pi^0}(E_{\gamma},\vec{r}) dr
\ee
The above calculation is for pion production from $pp$ interactions only.
The effect of the known chemical composition of the ISM can be taken into account by 
increasing the total pion production cross section by a factor of 1.30 
\citep{ms94}. 
\section{Reproducing the diffusive \gr emission from the Galactic center}
For our studies we reproduce the diffusive \gr emission calculating the line of sight integral
Eq.~\ref{grs:losi} for various $(l,b)$ combinations. As we are only interested in relative 
intensities, we can use instead of   Eq.~\ref{grs:losi} the relative emissivity
\be
\epsilon(l,b)&\propto&\int_{r=0}^{\infty}\rho(l,b,r)N_{CR}(l,b,r)\,dr
\label{calc:emissivity}
\ee
where $\rho(l,b,r)$ is the target material density as a function of
galactic coordinates and $N_{CR}(l,b,r)$ is the calculated CR
density at the point defined by triplet $(l,b,r)$.

\subsection{Gas distribution near the Galactic center}
The inner $\pm 150$\,pc region of the Galaxy contain interstellar $H_2$ gas
of about 2 to 5$\times10^7$ solar masses 
\citep{tsuboi99} 
in a rather complex setup of molecular clouds.
For our analysis we assumed that the target material density function can be
adequately described by the superposition of five spherical Gaussian
functions upon an asymmetric Gaussian base. These functions represent the 
molecular clouds associated with the radio arc of Sgr A, Sgr B, as well as the
longitude varying line-of-sight projection effect of the "Galactic Center Bow" as described
by \citet{tsuboi99}. Even though we are able to reproduce the observed line-of-sight
gas densities within about 5\% from the observed values, uncertainties in the exact depth 
distribution along the r-coordinate in Eq. \ref{calc:emissivity}, is expected
to result in a $\sim 50$\% systematic uncertainty in the final estimate of the diffusion coefficient.  
  
\subsection{Cosmic ray distribution near the Galactic center}
\begin{figure*}
\resizebox{\hsize}{!}{\includegraphics{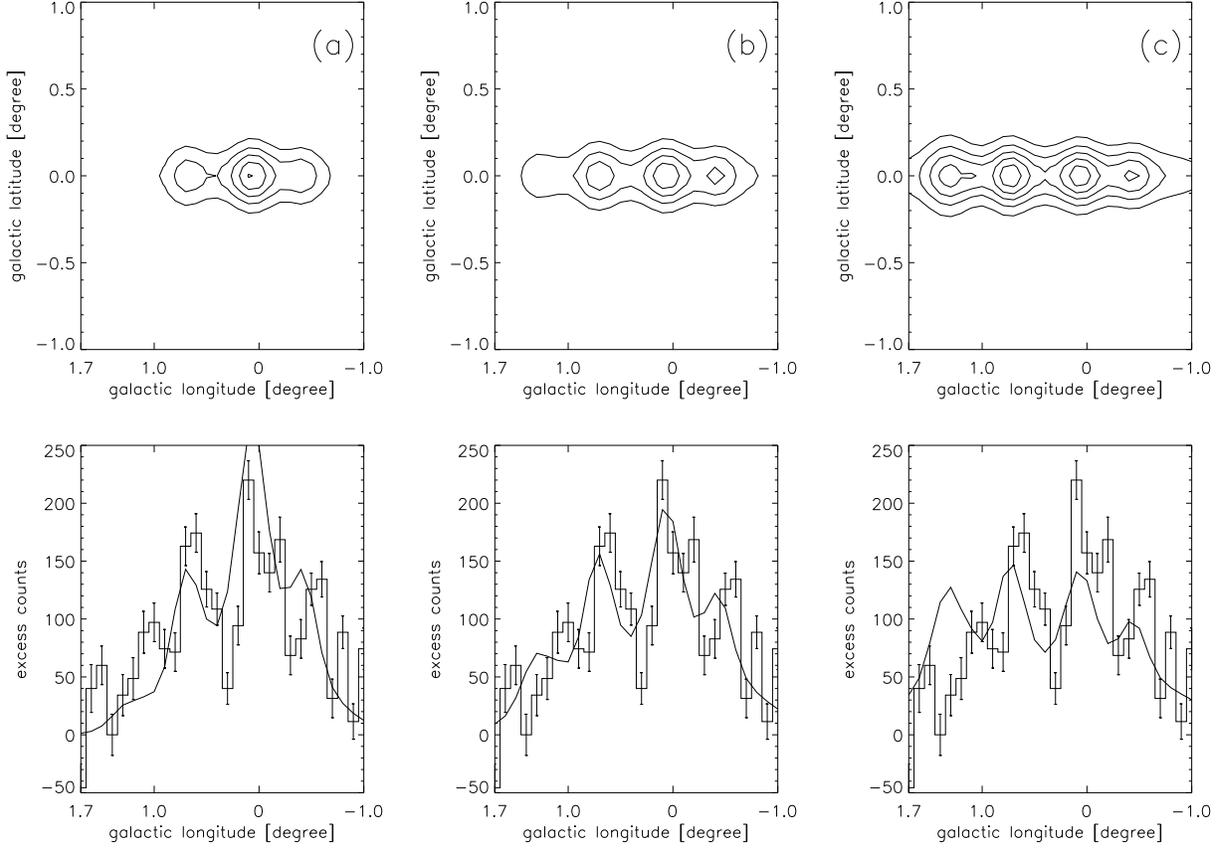}} 
\caption{
Calculated emission sky maps (top) together with the calculated excess
counts (bottom) shown as the solid lines.  The histogram indicate
the excess counts from the H.E.S.S. observation \citep{aharonian06}. 
The figures (a), (b) and (c) where generated for diffusion
coefficients with values $0.3\ kpc^2Myr^{-1}$, $1.3\ kpc^2Myr^{-1}$ (best fit)
and $15.1\ kpc^2Myr^{-1}$ respectively.}
\label{fig:res}
\end{figure*}

The CR distribution as a
function of spatial coordinates is calculated using 
the diffusive model of CR propagation.
We calculate the CR density assuming the CR 
coming from a single SNR event 10\,kyr ago, accelerating particles for a certain time. 
As was pointed out by \citet{maeda02}, the SNR Sgr A East has an estimated 
age of 10\,kyr, but also younger ages have been given for that SNR \citep{rockefeller05}.
For our study, we assume that Sgr A East has an age of 10\,kyr and was  
accelerating CR on various timescales less than 10\,kyr. 
We are  interested in the CR distribution near the source and thus neglect 
the effects of spatial inhomogeneities in the interstellar gas density (which are small for CR protons with the 
 energies we are dealing with) by assuming a mean gas density,
and also  boundary effects by imposing boundary conditions for the CR propagation problem at infinity. 
In this case one can find a solution for the propagation equation in the 
literature \citep{syrovatskii59}.
Using the integral in Eq.~\ref{calc:emissivity} the emission seen by H.E.S.S.
\citep{aharonian06} can be recreated for different diffusion
coefficients as shown in Fig.~\ref{fig:res} (top). As noted  by
\citet{aharonian06}, the excess counts observed by H.E.S.S.
follow the known target material density fairly well, except for the
region $l\ge1^\circ$. This provides a gauge for approximating the
diffusion coefficient governing the CR propagation near the galactic
center.

To obtain better counting statistics, we compare the model and observed
H.E.S.S. excess counts, integrated over Galactic latitude. 
The excess counts observed by H.E.S.S. then are
proportional to the integral
\be
\epsilon(l)&\propto&\int_{r=0}^{\infty}\int_{b=-\pi2}^{\pi/2}\rho(l,b,r)N_{CR}(l,b,r)r\,db\,dr,
\label{calc:int}
\ee
where the CR density $N_{CR}$ has been 
calculated for different diffusion coefficients, as shown in Fig.~\ref{fig:res}
(bottom). 
The obtained fit is normalised to the total numbers 
of excess counts before calculating the corresponding  $\chi^2$ value. 
The results of the analysis are shown
in Fig.~\ref{fig:chisq} below.  The optimal value for the diffusion
coefficient $k$ was found to be
\begin{equation}
\kappa=1.3\,{\rm kpc}^2{\rm Myr}^{-1},
\label{calc:kfinal}
\end{equation}
which is about 40\% smaller than the diffusion coefficient estimated by
\citet{aharonian06}. We also note that the minimum reduced $\chi^2$ is 1.5
(for $\sim 25$ d.o.f.), which indicates that the data are marginally
described by the model. Note that we do observe
a well-defined minimum in Fig.~\ref{fig:chisq}, leading to the abovementioned
measurement of $\kappa$. It would be interesting to see if more sophisticated
3D modelling of the total gas distribution result in even smaller $\chi^2$ values. Because of this uncertainty, we estimate a systematic uncertainty of $\sim 50\%$ on the measured value of $\kappa$.

\begin{figure}
\resizebox{\hsize}{!}{\includegraphics{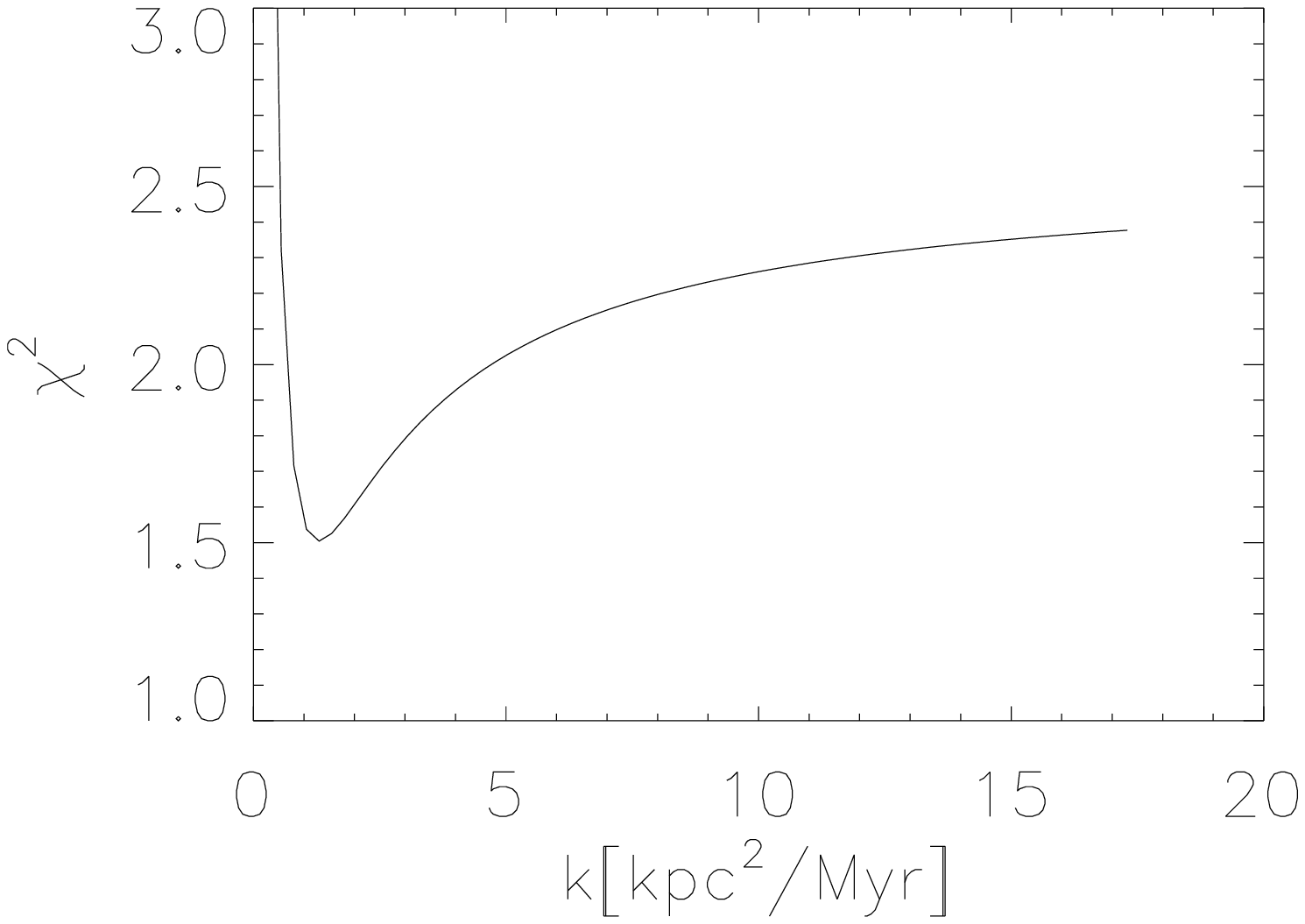}} 
\caption{
Reduced $\chi^2$ values plotted for different diffusion coefficients. We find a
well defined minimum for a diffusion coefficient of $k=1.3\ kpc^2Myr^{-1}$. 
}
\label{fig:chisq}
\end{figure}
\subsection{Mean CR energy}
In the last section, we have shown that the diffuse \gr emission from the 
Galcatic center ridge can be explained by CR hadrons which propagation can be described by a diffusion coefficient of 
$1.3$\,kpc$^2$Myr$^{-1}$.
This diffusion coefficient is valid for CR generating the bulk of
the \gr emission in the H.E.S.S. energy range. 

\begin{figure}
\resizebox{\hsize}{!}{\includegraphics{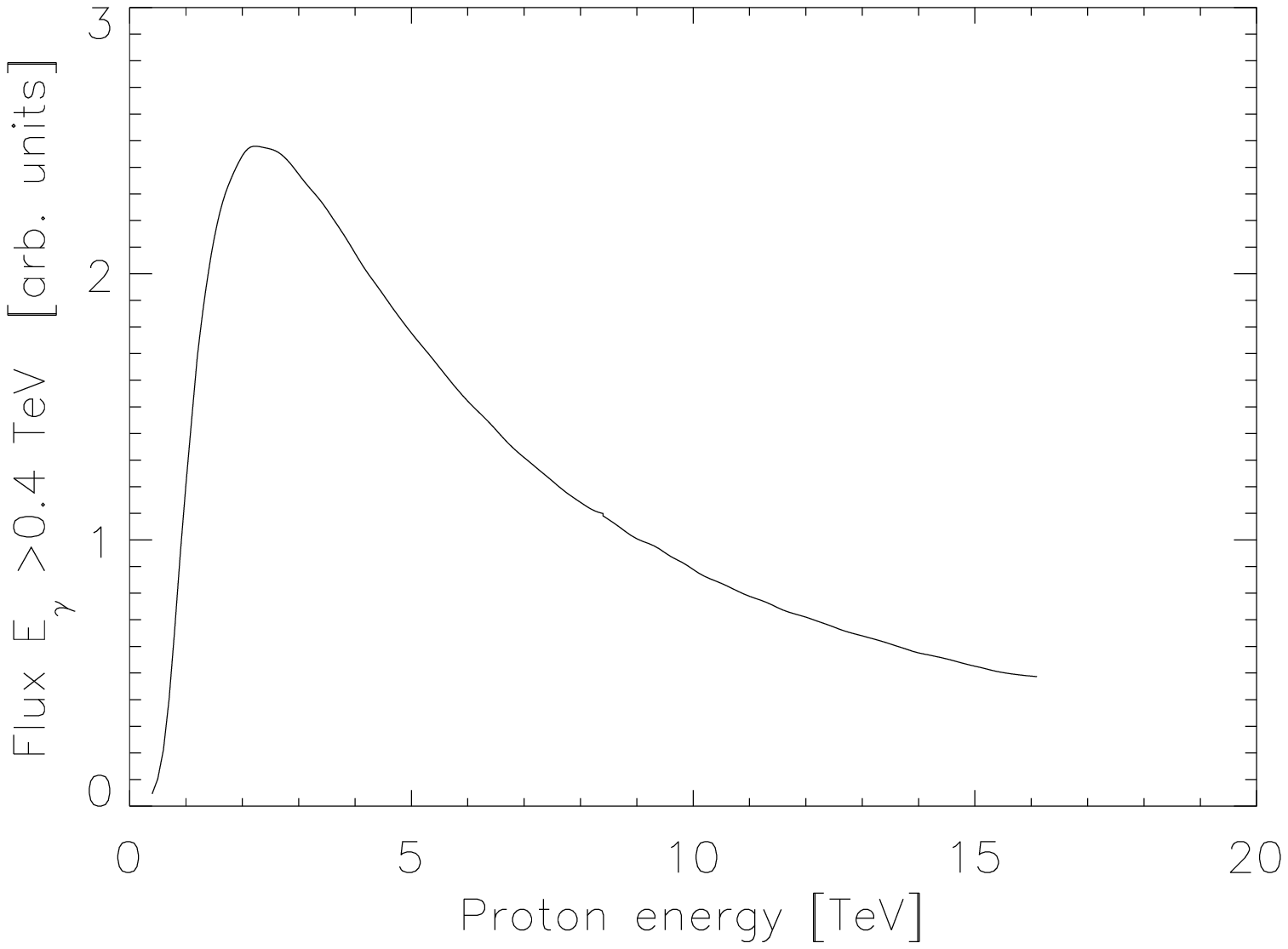}} 
\caption{
Number of photons with $E_{\gamma}>400\,GeV$ for different CR proton energies
assuming a CR proton spectral index of $s\,=\,2.29$.}
\label{fig:gl400}
\end{figure}

To compare this value 
with that obtained by fitting local CR data, as derived e.g. by \citet{moskalenko02},  \citet{jones01} or \citet{maurin02}, we have to  estimate the energy of the CR probed here.
For this investigation, we  approximate the sensitivity of the H.E.S.S. telescope array by a box function from 0.4\,TeV to 20\,TeV. For the CR proton spectrum 
we assumed a power law with the observed  photon index  $\Gamma\,=\,2.29\pm 0.27$ \citep{aharonian06}.
We used the Pythia \citep{pythia1,pythia2} event generator package to calculate the number 
of \grs with $E_{\gamma}>0.4$\,TeV, for different CR proton energies.
The result of this calculation is shown in Fig.~\ref{fig:gl400}.
The maximum number of photons is produced by CR protons with energies of about 2.2\,TeV, given a proton spectral index of 2.29. Note that the location of the maximum  
depends on the proton spectral index. Given the uncertainties stated by  \citet{aharonian06}, 
we find that CR protons with energies in the range 1.7\,TeV to 3\,TeV contribute 
mostly to the \grs seen by H.E.S.S.
Assuming a diffusion coefficient of the form 
\be
\kappa&=&\kappa_0\left(\frac{\xi}{\xi_0}\right)^{0.6},
\ee
where $\xi_0=1\,$GV and $\xi$ the particle rigidity
we find $\kappa_0=0.013$\,kpc$^2$Myr$^{-2}$,
which is significantly smaller than the values found  
by fitting local CR data. However, $\kappa$ for the latter range from 0.0535\,kpc$^2$Myr$^{-2}$ to 0.201\,kpc$^2$Myr$^{-2}$, as compiled by \citet{maurin02} (with references therein).
This finding can be well explained by enhanced turbulence and a higher field strength of the interstellar magnetic field in the  
Galactic center region. We note however that the rigidity dependence of 0.6 can only apply to a limited
energy (rigidity) range, since $\kappa$ must always be larger than the Bohm limit.

\section{Summary and Discussion}
We have shown that the progress in the imaging Cherenkov technique now makes it 
possible for the first time to measure the CR diffusion coefficient  
in other parts of the Galaxy. A diffusion coefficient of $\kappa\sim 1.3$ kpc$^2$Myr$^{-1}$
appears to be well measured from the data, although we have to add a $\sim 50\%$ systematic
undertainty arising from uncertainties in the actual 3D gas density distribution, as well
as uncertainties on the epoch when the central source activity started, which was assumed to be 10 kyr
in this paper. The most likely central source is Sgr A East, for which the epoch of onset of
activity is $\sim 10$ kyr, but if the central source was the central massive black hole Sgr A*,
the timescale of activity would be much less certain, resulting in even larger uncertainties
on $\kappa$. Note however that, given an initial epoch of onset of activity (i.e. 10 kyr),
estimates of $\kappa$ appear to be fortunately robust
against uncertainties in the detailed time profile of particle acceleration within 5 kyr after the SN explosion. For example, 
reducing the ``on''-time of the source by a factor of five, decreases the diffusion coefficient by 40\%. Such detailed studies will however be treated in a subsequent paper. 
Thus, if Sgr A East was the source of CR, then we have a relatively
accurate masurement of $\kappa$. 

In general, following the notion of \citet{aharonian06} to constrain the cosmic ray diffusion
coefficient $\kappa$ from the spatial distribution of diffuse $\gamma$-rays resulting
from impulsive injection of a CR source at some time in the past, we have
shown that one can obtain unique measurements of the diffusion coefficient,
provided that the gas density distribution, as well as the epoch of onset of
central source activity is known. 
We also showed that the CR diffusion coefficient in the Galactic center region is 
significantly smaller than that obtained by fitting local CR data. Our understanding
of turbulence theory is however still too limited to understand how diffusion coefficients
scale with the turbulence $\delta B$, the correlation length associated with this
turbulence, the total magnetic field strength $B$ and rigidity
dependence for perpendicular and parallel diffusion. Hopefully our new
measurement of $\kappa$ will help to constrain results from turbulence theory.

\end{document}